\documentclass{elsarticle}
 
\usepackage{lineno,hyperref}
\modulolinenumbers[5]
\usepackage{bm}
\usepackage{graphicx}
\usepackage{amssymb}
\usepackage{amsmath}
\usepackage{eufrak}
\usepackage{color}
\usepackage[dvipsnames]{xcolor}
\usepackage[utf8]{inputenc} 
\usepackage{hyperref}
\usepackage{pifont}
\usepackage{ulem}
\usepackage{array}
\usepackage{verbatim} 
\usepackage{placeins}

\newcommand{\nix}[1]{}

\begin{document} 
	
\title{Circular and Linear Photogalvanic Effects in Type-II GaSb/InAs Quantum Well Structures in the Inverted Regime}
	\author{H.~Plank$^1$, S. A. Tarasenko$^2$, T. Hummel$^1$, G. Knebl$^3$, P. Pfeffer$^3$, M. Kamp$^3$, S. H\"ofling$^{3,4}$, and S.\,D.~Ganichev$^1$}
	\address{$^1$ Terahertz Center, University of Regensburg, Regensburg, Germany}
	\address{$^2$ Ioffe Institute, St. Petersburg, Russia}
	\address{$^3$ Technische Physik 
		University of W\"urzburg, 
		W\"urzburg, Germany}
	\address{$^4$ University of St Andrews, St Andrews 
		United Kingdom}
\begin{abstract}	
We report on the observation of photogalvanic effects induced by terahertz radiation in type-II GaSb/InAs  quantum wells with inverted band order. 
Photocurrents are excited at oblique incidence of radiation and consists of several contributions varying differently with the change of the radiation polarization state; 
the one driven by the helicity and the other one driven by the linearly polarization of radiation are of comparable magnitudes.  
Experimental and theoretical analyses reveal that the photocurrent is dominated by the circular and linear photogalvanic effects in a system with a dominant structure inversion asymmetry. 
A microscopic theory developed in the framework of the Boltzmann equation of motion considers both photogalvanic effects and describes well all the experimental findings. 
\end{abstract}
	
	\maketitle{} 
	
\section{Introduction}		
In recent years, type-II GaSb/InAs two dimensional (2D)
structures have attracted growing attention in both theoretical and applied research. 
A distinguish feature of the GaSb/InAs quantum well (QW) structures  and superlattices is that  electrons and holes are mostly localized in the InAs and GaSb layers, respectively, 
and that the energy gap between the electron and hole subbands 
can be efficiently tuned by adjusting the widths of the layers, for reviews see e.g~\cite{Det_Kroemer,book_IR-Det,book_Zhang}.  
Narrow gap type-II GaSb/InAs superlattices, being characterized by a small effective mass, high mobility, excellent electron confinement, 
and the possibility of modulation doping, became an important system 
for mid-infrared radiation detection~\cite{Det_Kroemer,book_IR-Det,Det_Mohseni,Det_Wei,Det_Walther,Det_Ting,Det_Hoang} and were proposed as a 
novel candidate for terahertz (THz) radiation detectors~\cite{Li2008,Det_Li}.
Furthermore, it has been shown most recently that type-II GaSb/InAs QWs can be 
tuned through the topological quantum phase transition by the variation of the layer thicknesses and/or the application of front- and back-gate voltage~\cite{book_Zhang,ZhangLiu2008}. 
This property should enable the fabrication of high-quality materials with an 
interchanged band order,
a crucial issue for nontrivial 2D topological insulators (TIs)~\cite{HasanKane2010,Moore2010}.  
The inverted band order results in the emergence of topologically protected helical edge states, where 
carriers with opposite spin projections counter-propagate along each edge~\cite{book_Zhang,ZhangLiu2008} and supporting the spin quantum Hall effect~\cite{koenig,Knez2011b}.
Contemporary investigations of GaSb/InAs QWs in the topologically non-trivial phase  
comprise theoretical works on their electronic properties~\cite{ZhangLiu2008,Xu,Michetti,Klipstein,Qu,Papaj} and  transport measurements~\cite{koenig,Knez2011b,Knez2011,Wegscheider2013,Suzuki2013,Spanton2014,Wegscheider2014,Knez2014,Wegscheider2015,Du2015,Nyuyen2015,Suzuki2015,Ganjipour2015}.

Here, we report on the observation of a THz radiation induced linear 
and circular photogalvanic effects in type-II GaSb/InAs QWs in the inverted regime. 
Photogalvanic spectroscopy bridges electron transport and
optics and has become a very efficient tool to study nonequilibrium processes in semiconductors and low dimensional structures,  
yielding information on their point-group symmetry, 
details of the band spin splitting, processes of momentum, energy, and spin
relaxation etc., for reviews see~\cite{Ganichev2003,Ivchenko2008}. Most recently, the circular photogalvanic effect has been applied to study the Rashba/Dresselhaus spin-orbit coupling in  non-inverted type-II GaSb/InAs superlattices excited by polarized infrared radiation~\cite{GanichevGolub2014,Li2015}. Furthermore, the circular photogalvanic effect, resulting in a photon-helicity-dependent photocurrent,  is a key ingredient for the realization for a all-electric scheme to detect the radiation Stokes parameters~\cite{danilov2009}. 
Our experiments demonstrate that  photogalvanic currents can be excited efficiently
in type-II GaSb/InAs QWs in the frequency range from about 1 to 3.5 THz. Linearly or circularly polarized
terahertz radiation induces a $dc$ electric current whose magnitude and direction depend on the radiation polarization 
state. In particular, the photocurrent reverses its direction by switching the radiation polarization 
from left-handed to right-handed. We present a kinetic theory of the observed 
circular and linear photogalvanic effect which describes well the experimental data. 
We suggest that the photocurrents are of orbital origin and stem
from the lack of an inversion symmetry in the structure. 
Experimental data on the photocurrent anisotropy reveal that the dominated inversion symmetry breaking mechanism
responsible for the THz induced photocurrents is the structure 
inversion asymmetry. 

\section{Samples and technique}

\subsection{Samples}
\begin{figure*}
	\includegraphics[width=\linewidth]{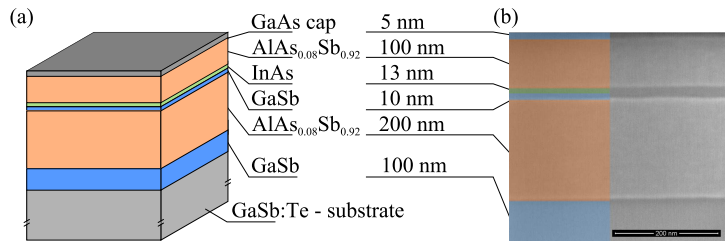}
	\caption{Heterostructure design, panel (a), and SEM image
	of the sample cross section, panel (b). 
	}
	\label{figure_1_sample}
\end{figure*}

The samples were fabricated on Te-doped (100)-oriented GaSb substrates by molecular beam epitaxy (MBE). 
The MBE chamber was equipped with evaporation cells for group III elements (Al, Ga, In) and with cracking cells for group V elements (As, Sb). 
The substrate temperature was controlled by pyrometer during growth. The wafer was heated to 300$^\circ$C pre-growth in the load lock chamber. Oxide desorption was realized at 580$^\circ$C under Sb stabilization flux. 
Afterwards the growth was started with a 100 nm thick GaSb smoothing layer followed by the lower 200 nm lattice matched AlAs$_{0.08}$Sb$_{0.92}$ barrier. The explicit compound quantum well consists nominally of a 10 nm GaSb and a 13 nm InAs well. On top a 100 nm AlAs$_{0.08}$Sb$_{0.92}$ barrier and a 5 nm GaSb cap finish the structure. 
The composition of the QW widths aims at an inversion of the electron and hole subbands. 8x8 kp simulations suggest an inversion of about 70~meV at $k=0$.
Scanning electron microscopy (SEM) images of the samples cross section, structure composition and thickness of all layers are shown in Fig.~\ref{figure_1_sample}.  To probe the photocurrent in different directions four pairs of ohmic contacts were prepared; two pairs in the corners of the $5\times5$\,mm$^2$ squared samples and two in the middle of the edges, see Fig.~\ref{figure_2_phi}. The samples' edges are oriented along the $[110]$ and $[0{\bar1}1]$ directions.  
To characterize the samples, electrical measurements were done 
%
on Hall-bar structured samples 
at helium temperature. In these measurements the mobility is $\mu = 6.9\times10^{4}$~cm$^2$/Vs and the electron density is $n_e= 1.2\times 10^{11}$~cm$^{-2}$. 
Transport measurements show the coexistence of $n$- and $p$-type carriers and support the assumption of the sample being in the inverted regime. 
%

\subsection{Technique}
To excite photocurrents we applied a high power line-tunable  NH$_3$ THz laser~\cite{Lechner2009,Drexler2013,ratchet2009}
optically pumped by a pulsed transversely excited atmosphere (TEA) CO$_2$ laser~\cite{SGEopt2003,Chongyun}. 
The laser operates at the frequencies $f$ = 1.07, 2.03, and 3.31\,THz. The corresponding photon energies ($\hbar \omega = 4.4$, 8.4, and 13.7\,meV, respectively) are smaller than the band gap as well as the size-quantized subband separation. The radiation induces indirect optical transitions (Drude-like free-carrier absorption) in the lowest conduction subband. The laser generates single pulses with a duration of about 100\,ns and a repetition rate of 1\,Hz.
The laser peak power, being of the order of $P\approx10$~kW, was controlled by the THz photon drag detector~\cite{book}. 
The laser radiation was focused onto a spot size of about 1.6 - 2\,mm diameter depending on the radiation frequency. 
The beam had an almost Gaussian profile which was measured by a pyroelectric camera\,\cite{Ganichev1999,Ziemann2000}.
Various polarization states of the radiation are achieved by
transmitting the linearly polarized ($\bm E \parallel y$) laser beam through $\lambda$/4 or $\lambda$/2 crystal quartz plates. 
By rotating a $\lambda$/4 plate one transfers the
linear into elliptical polarization. The polarization states are directly related to the angle
$\varphi$ between the initial linear polarization of the laser light  and the optical axis of the plate.
By that we obtain the degree of circular polarization given by $P_{\rm circ} = \sin{2 \varphi}$ and the bilinear combinations of the linear polarization vector components. The latter following the Stokes parameters~\cite{Stokes1,Stokes2} are  expressed via $\sin 4\varphi$ and $\cos 4\varphi$~\cite{BelkovSSTlateral,ratchet2011}.   	
In some experiments we also used $\lambda$/2 quartz plates providing a linearly polarized radiation 
with the polarization plane rotated by the azimuth angle $\alpha$ from the initial linear polarization of the laser light along the $y$ axis.
The experimental geometry is sketched in the insets in panel (c) of Fig.~\ref{figure_2_phi}. 
Unbiased samples were excited at oblique incidence with the plane of incidence lying in the ($xz$) plane. The angle of incidence $\theta_0$ was varied between $-30^\circ$ to +30$^\circ$, with respect to the layer normal $z \parallel [001]$. 
Both photocurrent components perpendicular, $J_y$, and parallel, $J_x$, to the plane of incidence were investigated. 
In experiments aimed at the study of photocurrent anisotropy, we also used other orientations of the plane of incidence
with respect to the $x$ and $y$ axis. Details of these geometries are discussed below. To avoid any signals stemming from 
the illumination of contacts or the sample edges, the samples were excited through a
2.7~mm diameter aperture in a metal mask covering the contacts and the edges. The photocurrents were measured by the 
voltage drops across a 50\,$\Omega$ resistor, Fig.~\ref{figure_2_phi}(c). The peak voltage was recorded with
a 1~GHz bandwidth storage oscilloscope. Several samples of the same batch were investigated at room temperature 
yielding the same results.

\section{Results}
%
\begin{figure*}
	\includegraphics[width=\linewidth]{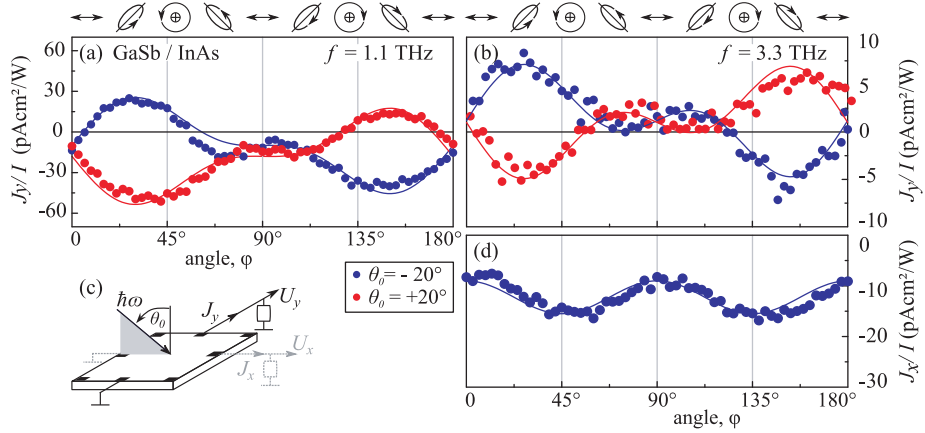}
	\caption{Dependence of the normalized photocurrent $J/I$ 
	on the angle $\varphi$  measured in (001)-grown type-II GaSb/InAs QW  at room temperature. 
	Here $I$ is the radiation intensity. The data are shown for the ($xz$)-plane of incidence and the  angle of incidence $\theta_0 = \pm 20^\circ$, see panel (c). Panels (a) and (b) show the photocurrent $J_y / I$ excited in the direction perpendicular to the plain of incidence
	by the radiation with frequency $f= 1.1$ and 3.3~THz, respectively. Panel (d) shows the  data for the longitudinal photocurrent $J_x$.  Solid line are fits after Eqs.~(\ref{j_final}). Arrows, ellipses and  circles 
	on the top illustrate the state of polarization for several values of the angle $\varphi$. 
	}
	\label{figure_2_phi}
\end{figure*}
\begin{figure*}
	\includegraphics[width=\linewidth]{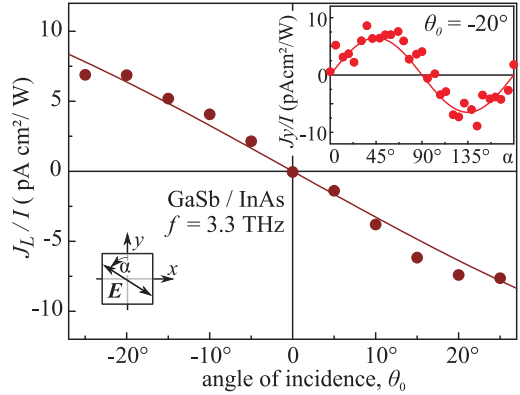}
	\caption{Amplitude of the longitudinal 
		photocurrent excited by linearly polarized radiation as a function of the angle of incidence. Top right 
		inset shows the dependence of the normalized photocurrent $J_y/I$ on the azimuth angle $\alpha$  measured at room temperature. 
		The data are shown for the ($xz$)-plane of incidence and the  angle of incidence $\theta_0 = - 20^\circ$. 
		Bottom left inset defines the azimuth angle $\alpha$. Solid lines are fits after Eqs.~(\ref{j_final}) 
		re-written for the case of linear polarized radiation, see discussion after Eqs.~(\ref{j_final}).
	}
	\label{figure_3_alpha}
\end{figure*}

Irradiating the middle of (001)-grown type-II GaSb/InAs QW (edges are not illuminated) with polarized THz radiation at oblique incidence photocurrent we detect a signal in the absence of an external bias. The signal follows the temporal structure of the laser pulse. Its response time is determined by the time resolution of our setup but it is at least 1~ns or shorter. 
This fast response is typical for photogalvanic effects where the signal decay time is expected to be of the order of the momentum relaxation time~\cite{SturmanFridkin,Ivchenkobook2,book}, which according  to transport data in our samples at room temperature is much shorter 1~ns.
The photocurrent is observed in the investigated frequency range from 1.1 and 3.3~THz. 
Photosignals are observed both in the direction perpendicular
to the plane of radiation incidence [transverse geometry, $J_y$ in Figs.~\ref{figure_2_phi}(a) and (b)]
and in the direction lying in the incidence plane 
[longitudinal geometry, $J_x$ in Fig.~\ref{figure_2_phi}(d)].
Figures~\ref{figure_2_phi} shows the dependences of the photocurrent components $J_y$ and $J_x$ on the angle $\varphi$ for the angles of incidence  $\theta_0 = \pm 20^\circ$. 
We find that the polarization dependence of the transverse current is well fitted by 
$J_y =  J_{\rm C} \sin{2 \varphi} + J_{\rm L}  \sin{4 \varphi}\,/2 + J_{\rm off}$ and that of the 
longitudinal current by $J_x = - J_{\rm L} (1- \cos{4 \varphi})/2 + J^\prime_{\rm off}$, where $J_{\rm C}, J_{\rm L}$ and $J_{\rm off}$ are fitting parameters. 
A distinguish feature of the photocurrent contribution $J_2 \sin{2 \varphi}$ is that 
it reverses the direction upon switching the radiation polarization from right-handed 
($\varphi = 45^\circ$) to left-handed ($\varphi = 135^\circ$) circular
polarization. Note that both contributions, $J_{\rm L}  \sin{4 \varphi}/2$ and $J_{\rm L} (1- \cos{4 \varphi})/2$, are equal to zero for circularly polarized radiation. These contributions are driven by the linear polarization component of elliptically polarized radiation.
To prove this the photocurrent was measured with linearly polarized radiation. 
The top right inset in Fig.~\ref{figure_3_alpha} shows an exemplary dependence 
of the transverse photocurrent $J_y$ on the azimuth angle $\alpha$ obtained at the incidence angle $\theta_0 = - 20^\circ$.
The polarization dependencies of the transverse and longitudinal (not shown) components of the photocurrents
are well described by $J_y = J_{\rm L} \sin{2 \alpha} + J_{\rm off}$ and $J_x = - J_{\rm L} (1 - \cos{2 \alpha}) + J^\prime_{\rm off}$, respectively, using $J_{\rm L}$ and $J_{\rm off}$ obtained in the experiments with elliptically polarized radiation.
Note that the offsets $J_{\rm off}$ and $J^\prime_{\rm off}$ have magnitudes substantially smaller
than the amplitudes of the photocurrents $J_{\rm C}$ and $J_{\rm L}$ and are out of scope of our paper.
%

\begin{figure*}
	\includegraphics[width=\linewidth]{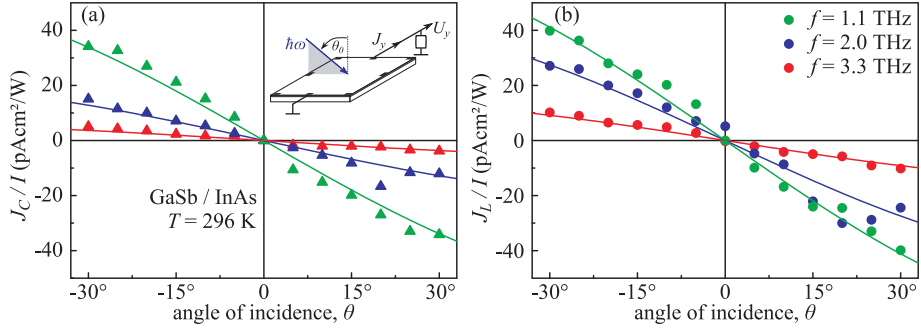}
	\caption{
	Dependencies of the normalized circular,   $J_{\rm C}/I$, and linear,   $J_{\rm L}/I$,  photocurrent contributions
	on the angle of incidence $\theta_0$  measured for three radiation frequencies. 
	Inset shows the experimental geometry. Solid lines are fits after Eqs.~\ref{j_final}.
	}
	\label{figure_4_theta}
\end{figure*}

As a function of the incidence angle all photocurrent contributions change signs at $\theta_0 \approx 0$. This is demonstrated 
for photocurrent components $J_{\rm C}$ and $J_{\rm L}$ in Fig.~\ref{figure_4_theta}. The similar behaviour is also observed for the transverse and longitudinal photocurrents induced by linearly polarized radiation, see Fig.~\ref{figure_3_alpha} showing the dependence of the transverse photocurrent on the angle of incidence.
Figure~\ref{figure_4_theta} presents a set of the experimental data on the circular and 
linear photogalvanic currents obtained for THz radiation of different frequencies. 
These data reveal that the increase of the radiation frequency leads to the decrease of the magnitudes of both photocurrent
contributions. Such a trend corresponds to the well known spectral behavior of the Drude absorption~\cite{Seeger04} at 
$\omega \tau_p \geq 1$, where $\tau_p$ is the momentum relaxation time of carriers.  
Below we demonstrate that the observed polarization dependences, the variation of the photocurrent with the angle of incidence, as well as the frequency behavior follows the microscopic theory of the circular and linear photogalvanic effects at the Drude-like absorption in (001)-oriented QWs.

The observed circular photogalvanic effect (the photocurrent proportional to $J_{\rm C}$) allows us to analyse the relative strength and interplay of the structure inversion asymmetry (SIA) and the bulk inversion asymmetry (BIA) of the studied samples.
The technique relies on the phenomenological linear coupling between the polar vector of the photocurrent and the axial vector of the photon angular momentum, which is a prerequisite for the circular photogalvanic effect to occur and  
which can originate in QWs from SIA or BIA, see also Refs.~\cite{GanichevGolub2014,Li2015,Giglberger2007,Kohda2012}.    
To judge on the dominant mechanism of inversion symmetry breaking causing the photocurrent generation, 
we have studied the photocurrent anisotropy in the QW plane. 
To this end we have measured the longitudinal and transverse circular photocurrents for four different orientations of the radiation incidence plane: ($xz$), ($yz$), and the planes ($x^\prime z$) and ($y^\prime z$) containing the cubic axes $x^\prime \parallel [100]$ and $y^\prime \parallel [010]$. 
Our measurements reveal that in all these geometries the circular photocurrent is reliably detected in the direction perpendicular to the incidence plane only and being almost of the same magnitude. 
These findings correspond to the photocurrent caused by SIA. The circular photogalvanic current caused by BIA would flow in the plane of radiation incidence (longitudinal photocurrent) for the incidence planes ($x^\prime z$) and ($y^\prime z$). 
Moreover, it would have an opposite sign of the transverse components for the incidence planes ($xz$) and ($yz$). Therefore,  together with the SIA contribution we would observe substantial difference in the magnitude. Both features are not detected. Thus, we conclude that the dominant mechanism of the photocurrent generation in the samples under study is related to SIA.

Finally, we note that illuminating the edges of the sample without a metal mask we have observed a strong circular photocurrent. 
The photocurrent is excited at normal incidence. This observation is in contrast to the data discussed above, where the $dc$ current for $\theta = 0$ vanishes.  
As our samples are in the inverted regime and the photon energy of the THz radiation used is rather small, this is probably an evidence that the topological helical edge states are involved in the formation of such a photon helicity driven photocurrent. However, this edge photocurrent requires a further careful study and is out of scope of the present paper. 
%
\section{Microscopic model}

In our experiments the photon energy of the THz radiation $\hbar\omega$ is much smaller than the electron Fermi energy $E_F$ (a few hundreds meV). Therefore, a relevant approach for the theoretical description of the photocurrent generation is quasi-classical. In this approach~\cite{Tarasenko2011,Budkin2016} the THz radiation is considered as an $ac$ electric field 
\begin{equation}
\bm{E}(t) = \bm{E} e^{-i \omega t}+\bm{E}^* e^{i \omega t} \:,
\end{equation}
where $\bm{E}$ is the field amplitude, which acts upon the electrons. The in-plane component $\bm{E}_{\parallel}(t)$ of the field drives the electrons back and forth in the QW plane inducing an $ac$ electric current at the field frequency $\omega$. The $ac$ current does not follow the time dependence $\bm{E}_{\parallel}(t)$ but is retarded with respect to the in-plane field by the phase $\arctan(\omega\tau_p)$, where $\tau_p$ is the momentum relaxation time. The out-of-plane component $E_z(t)$ pushes the electron density in the QW normal to the $z$ or $-z$ directions depending on the field polarity. This shift of the electron density along the growth direction in strongly asymmetric systems such as GaSb/InAs heterostructures results in a change of the electron mobility. The modulation of the electron mobility with the same frequency $\omega$ results in a rectification of the $ac$ electric current, which causes the emergence of a $dc$ component of the current. For clockwise and counter clockwise rotating electric fields, corresponding to the $\sigma^+$ and $\sigma^-$ circular polarized radiations, respectively, the $dc$ current flows in the opposite directions.

The rate of electron scattering, which determines the mobility, can be presented in general in an asymmetric heterostructure up to the first order in $E_z(t)$ in the form 
\begin{equation}\label{W_pp}
W_{\bm{p} \bm{p}'} = W_{\bm{p} \bm{p}'}^{(0)} + e E_z(t) \: W_{\bm{p} \bm{p}'}^{(1)} \:,
\end{equation}
where $W_{\bm{p} \bm{p}'}^{(0)}$ is the scattering rate at zero field, $W_{\bm{p} \bm{p}'}^{(1)}$ is the field-induced correction related to SIA and $e$ is the electron charge. For such a scattering rate, the kinetic theory developed in Refs.~\cite{Tarasenko2011,Budkin2016}
yields the following expression for the $dc$ electric current density 
%
%
\begin{equation}\label{j}
\bm{j} = - \frac{e^3 \zeta_1 N_e \tau_p^2}{m^*}  
\left( \frac{\bm{E}_{\|} E_z^*}{1 - i \omega\tau_p} + \frac{\bm{E}_{\|}^* E_z}{1 + i \omega\tau_p} \right) \,,
\end{equation}
where $N_e$ is the electron density, $m^* = p_F / v_F$ is the effective mass at the Fermi level, $p_F$ and $v_F$ are the Fermi momentum and Fermi velocity, respectively, $\tau_p$ is the momentum relaxation time,
\begin{equation}
\tau_p^{-1} = \sum_{\bm{p}'} W_{\bm{p} \bm{p}'}^{(0)}\left[1-\cos (\varphi_{\bm p} - \varphi_{\bm p'} )\right] \:,
\end{equation}
$\zeta_1$ is the parameter of the scattering asymmetry,
\begin{equation}
\zeta_1 = \sum_{\bm{p}'} W_{\bm{p} \bm{p}'}^{(1)}\left[1-\cos (\varphi_{\bm p} - \varphi_{\bm p'} )\right] \:,
\end{equation}
and $\varphi_{\bm p}$ and $\varphi_{\bm p'}$ are the polar angles of the vectors $\bm p$ and $\bm p'$ before and after the scattering event, respectively.
The current~(\ref{j}) is proportional to the square of the electric field amplitude, i.e. to the radiation intensity, and comprises both linear and circular photocurrents. Note that while in the theoretical consideration the current density ${\bm j}$ is used, in the experiments the electric current ${\bm J}$ is measured, which is proportional to the current density ${\bm j}$.

For the experimental geometry sketched in Fig.~\ref{figure_2_phi}(c), the components of the photocurrent~(\ref{j}) can be presented in the form
\begin{align}\label{j_final}
j_{x} = - \frac{e^3 N_e }{m^*}  \frac{ \zeta_1 \tau_p^2 E_0^2}{1+(\omega\tau_p)^2} \, t_p^2 \, \sin \theta \cos\theta 
[(1-\cos 4\varphi) /2] \,, \nonumber \\
j_{y} =  \frac{e^3 N_e }{m^*}  \frac{ \zeta_1 \tau_p^2 E_0^2}{1+(\omega\tau_p)^2} \, t_p t_s \, \sin \theta [ \sin 4\varphi /2 +  
\omega\tau_p \sin 2\varphi] \,,
\end{align}
where $t_p$ and $t_s$ are the amplitude transmission coefficients for $p$- and $s$-polarized radiation, respectively, 
\[
t_p = \frac{2 \cos{\theta_0}}{n_{\omega}\cos{\theta_0} + \cos{\theta}} \:, \;
t_s = \frac{2 \cos{\theta_0}}{\cos{\theta_0} + n_{\omega} \cos{\theta}} \:,
\]
$E_0$ is the amplitude of the electric field of the incident radiation, $\theta$ is the angle of refraction related to the incidence angle $\theta_0$ by $\sin \theta = \sin \theta_0 / n_{\omega}$, and $n_{\omega}$ is the refractive index. The component of the photocurrent in the incidence plane $j_x$ contains only a linear photogalvanic current while the perpendicular component $j_y$ contains both linear and circular photogalvanic currents. 

Curves in Figs.~\ref{figure_2_phi} (a), (b) and (d) show the polarization dependences of $J_x$ and $J_y$, respectively, calculated after Eqs.~(\ref{j_final}). 
They correspond well to these experimental data apart a small polarization-independent offset which is detected for some frequencies. 
The theory also describes well the polarization behavior of the photocurrent excited by linearly polarized radiation: In this case, the terms in the square brackets should be replaced by $\sin 2\alpha$ and $\cos 2\alpha$ for $J_x$ and $J_y$, respectively. 
This transformation just reflects a change of the Stokes parameters of radiation transmitted through $\lambda/2$ plate as compared to those transmitted through $\lambda/4$ plate. 
The calculated dependence of the photocurrent on the azimuth angle $\alpha$ is shown in Fig.~\ref{figure_3_alpha} for the transverse photocurrent excited by radiation with the frequency $f = 3.3$\,THz. 
We note that, in accordance with the Eqs.~(\ref{j_final}), for all fits in Figs.~\ref{figure_2_phi} and \ref{figure_3_alpha} we used the same prefactor for the linear photogalvanic effects excited at a given frequency. 
Equations~(\ref{j_final}) also reveal that the photocurrent emerges at oblique incidence only and flows in the opposite directions for
positive and negative angles of incidence $\theta_0$. 
Moreover, the equations explain the overall variation of the photocurrent with the angle of incidence. The corresponding calculated dependencies are shown by solid lines in Figs.~\ref{figure_4_theta} (a) and (b) for the circular and linear photogalvanic currents, respectively.

\section{Summary}
To summarize, our experiments show the photon helicity driven photogalvanic currents and photocurrents induced by the linearly polarized radiation can be effectively generated in type-II GaSb/InAs QWs at room temperature.
A microscopic theory developed in the framework of the Boltzmann equation of motion suggests that the photocurrents are of orbital origin and stem from the lack of an inversion symmetry in the structure. 
The performed experimental analysis of the circular photogalvanic effect anisotropy indicates that, in the context of photocurrents, the structure inversion asymmetry outweighs the bulk inversion asymmetry in the studied systems. 
In addition to photocurrents excited in the ''bulk'' of the QWs, we also observed a strong circular photogalvanic current stemming from the illumination of the sample edges. 
The edge photocurrent may be related to helical edges states in the studied QW structures characterized by the inverted band order, which is subject of a future work.  

The photocurrents determined by the radiation Stokes parameters, in particular, the one proportional to the radiation helicity, are detected in the whole investigated frequency range.  
This observation, together with a subnanosecond response time, suggests type-II GaSb/InAs QWs as a promising candidate for the fast room-temperature all-electric detectors of THz radiation polarization.

\section*{Acknowledgments}
We thank V.V. Bel'kov for fruitful discussions. The work was supported by the Elite Network of Bavaria (K-NW-2013-247), the DFG priority program SPP1666, the Volkswagen Stiftung Program, the State of Bavaria and the German Research Foundation (Ka2318/4-1). S.A.T. acknowledges support from the RFBR (projects 14-22-02102 and 16-02-00375).

\end{document}